\documentclass[twocolumn,showpacs]{revtex4}
\usepackage{times,xspace}
\usepackage{amsbsy,amssymb,amsmath,bm}
\usepackage{graphicx,color,epsfig,rotate}
\usepackage{fancyhdr}

\def\bbbc{{\mathchoice {\setbox0=\hbox{$\displaystyle\rm C$}\hbox{\hbox
to0pt{\kern0.4\wd0\vrule height0.9\ht0\hss}\box0}}
{\setbox0=\hbox{$\textstyle\rm C$}\hbox{\hbox
to0pt{\kern0.4\wd0\vrule height0.9\ht0\hss}\box0}}
{\setbox0=\hbox{$\scriptstyle\rm C$}\hbox{\hbox
to0pt{\kern0.4\wd0\vrule height0.9\ht0\hss}\box0}}
{\setbox0=\hbox{$\scriptscriptstyle\rm C$}\hbox{\hbox
to0pt{\kern0.4\wd0\vrule height0.9\ht0\hss}\box0}}}}

\newcommand{\ignore}[1]{}
\newcommand{\mComment}[1]{}
\newcommand{\gComment}[1]{}
\newcommand{\jComment}[1]{}
\newcommand{\rComment}[1]{}
\newcommand{\lComment}[1]{}

\renewcommand{\mComment}[1]{\textcolor{blue}{Manny: #1}}
\renewcommand{\gComment}[1]{\textcolor{red}{Gerardo: #1}}
\renewcommand{\jComment}[1]{\textcolor{green}{Jim: #1}}
\renewcommand{\rComment}[1]{\textcolor{magenta}{Ray: #1}}
\renewcommand{\lComment}[1]{\textcolor{purple}{Rolando: #1}}

\begin{document}

\title{Exact ground states of extended
$t-J_{z}$ models on a square lattice} 
\author{Zohar Nussinov}
\affiliation{Department of Physics, Washington University, St. Louis, MO 
63160, USA}
\author{Anders Rosengren}
\affiliation{Condensed Matter Theory,
KTH Physics,
AlbaNova University Center,
SE-106 91 Stockholm, Sweden}

\date{Received \today }

\begin{abstract}
We examine special extended 
spin $S=1/2$ fermionic and hard core 
bosonic $t-J_{z}$ models with nearest neighbor
and next nearest neighbor interactions to find 
exact ground states. Some of these
models display an exponentially large 
degeneracy with 
diagonal stripe like patterns. 

\end{abstract}


\maketitle

\section{Introduction} 

It has been nearly two decades since the discovery of the high-\(
T_{c} \) superconductors \cite{BM86}.  Although much has been learned
since, there is still no satisfactory explanation of what causes the
superconductivity. There is widespread agreement that it should be
possible to describe the electronic properties of the square CuO\(
_{2} \) lattices by the two-dimensional repulsive Hubbard model. This
well known model is given by

\begin{equation}
\label{Hubbard model}
H=-t\sum _{\langle ij\rangle \sigma }(c_{i\sigma }^{\dag }c_{j\sigma
}+c_{j\sigma }^{\dag }c_{i\sigma })+U\sum _{i} n_{i\uparrow }n_{i\downarrow},
\end{equation}%
where \( c_{i\sigma }^{\dag } \) creates an electron on site \( i \)
with spin \( \sigma \) and \( j \) is a nearest neighbor of \( i \).
This model contains both the movement of the electrons (hopping) (\( t
\), kinetic energy) and the interactions of the electrons if they are
on the same site (\( U \), potential energy).  The Hubbard model is
one of the simplest possible models of interacting electrons.  In the
large $U$ limit, the model reduces to the well known $t-J$ model
\begin{eqnarray}
H =
- t \sum_{\langle i j \rangle, \sigma}  (\overline{c}_{i, \sigma}^{\dagger} 
\overline{c}_{j, \sigma}
+ h.c.) +  J \sum_{\langle i j \rangle}  \vec{s}_{i} \cdot \vec{s}_{j}
\end{eqnarray}
 and the now constrained fermion operators
$\overline{c}_{i,\sigma}^{\dagger},\overline{c}_{i,\sigma}$ obey the
constraint of no double occupancy at any site $i$ ($n_{i} =
\sum_{\sigma}
\overline{c}_{i,\sigma}^{\dagger}\overline{c}_{i,\sigma}$ has
expectation values 0 or 1). In formal terms, the Fock space is
projected to this sector adhering to these
constraints. An even simpler version is afforded by the so-called
$t-J_{z}$ model given by
\begin{eqnarray}
H = - t 
\sum_{\langle i j \rangle, \sigma}  (\overline{c}_{i, \sigma}^{\dagger} 
\overline{c}_{j, \sigma}
+ h.c.) +  J_{z} \sum_{\langle i j \rangle}  s^{z}_{i} s^{z}_{j}
\label{tjzHam}
\end{eqnarray}
wherein the $SU(2)$ spin symmetry of the original Hubbard and $t-J$
models has been lifted. 

Numerical calculations on such models with open and cylindrical
boundary conditions have found non-uniform ``stripe'' patterns
\cite{WS} which have earlier been predicted by approximate solutions
to the Hubbard model \cite{jan}, and derived, very convincingly, by
the addition of the strong Coulomb effects present in the cuprates
\cite{steve}.  At the moment, there is no clear consensus as to
whether these patterns display the genuine ground state of the bare
Hubbard (or $t-J$) Hamiltonian or amount to a finite-size artifact.
Attacking these models has proven not to be an easy task. The Hilbert
space of the model is too large for exact numerical solutions.  Monte
Carlo simulations suffer from the sign problem. The interaction energy
\( U \) is approximately equal to \( 8t \), so we are neither in an
extremely strong, nor in a weakly interacting limit. As has been
emphasized by \cite{steve}, Coulomb repulsions in the cuprates are
poorly screened and universally promote the existence of stripes.
Notwithstanding, it would be of interest to attain rigorous 
results on whether various inhomogeneous electronic patterns 
can, in principle, be realized in the presence of 
short-range interactions and hopping alone in some models.

The less ambitious goal of this article is not directly related to the
very challenging endeavor of understanding such complicated systems
(albeit motivated by it).  Rather, we construct exactly solvable high
dimensional models related to the original $t-J_{z}$ models. 
Inspired by exactly solvable AKLT \cite{Assa, AKLT}
and in particular Klein models \cite{Klein82}, 
\cite{Chayes89}, \cite{BT}, \cite{NBNT}, \cite{Nussinov06}, we
construct an exactly solvable 
model whose individual pair building
blocks are an extension (containing superconducting pair terms and
additional phases for the fermionic version) of the $t-J_{z}$
approximation of the Hubbard model. The price for attaining this
solvable short range model 
(all interactions involve sites which are, at most,
two lattice constants apart) and others
is that their results cannot be directly related to the
models of more physical interest (such as the plain $t-J_{z}$ model).

\section{Construction of Quasi-Exact Solvable $S=1/2$ fermionic
and hard core bosonic systems on a square lattice} 

In the up and coming, we follow a generalization of the
Klein models \cite{Klein82}, \cite{Chayes89}, \cite{BT}, 
\cite{NBNT}, \cite{Nussinov06} yet now for fermionic $S=1/2$
particles on the square lattice. To do so, we first find, much in the
spirit of \cite{Klein82}, \cite{Chayes89}, \cite{BT}, 
\cite{NBNT}, \cite{Nussinov06} a 
quasi-exactly solvable $S=1$ model on a square
lattice \cite{explain_qe}. We then apply a Jordan 
Wigner transformation to convert the
$S=1$ problem into a fermionic $S=1/2$ system.  The resulting $S=1/2$
system is none other than a $t-J_{z}$ model augmented by
superconducting pairing terms and phase factors.

\subsection{Construction of ground states for the $S=1$ system}

First, we generalize the Klein model
(\cite{Klein82}, \cite{Chayes89}, \cite{BT}, \cite{NBNT}, \cite{Nussinov06}) 
constructions to the problem of $S=1$ spins
on the square lattice whose sites are labeled by $i$ (the locations
are $\vec{i} = (i_{x}, i_{y})$ are vectors with $i_{x}$ and $i_{y}$
integers). We assume that the lattice has periodic boundary conditions.
For clarity, we dispense with the vector sign henceforth. We
consider a spin $S=1$ system whose Hamiltonian can be written as a sum
over all plaquettes of the lattice
\begin{eqnarray}
H = \sum_{\Box} h_{\Box},
\label{h.}
\end{eqnarray}
with 
$h_{\Box}$ a polynomial of the total spin
of the four sites forming a given plaquette 
\begin{eqnarray}
\vec{S}_{\Box} = \sum_{i \in \Box} \vec{S}_{i},
\end{eqnarray}
i.e. 
\begin{eqnarray}
h_{\Box} = \mathcal{P}(\vec{S}_{\Box}^{2}) = \sum_{n=0}^{m} a_{n}
\vec{S}_{\Box}^{2n}.
\label{hp.}
\end{eqnarray}
Numerous polynomials (and hence short-range Hamiltonians) can be
engineered to find exactly solvable models. Before doing so let us
review the possible four and two particle
states of $S=1$ spins. With four spins in tow, the total spin of any
plaquette, $0 \le S_{\Box} \le 4$. The
Hilbert space spanned by four $S=1$ spins can be 
written as 
\begin{eqnarray}
1 \otimes 1 \otimes 1 \otimes 1 = \nonumber
\\ 4 \oplus (3)^{3} \oplus 
(2)^{6} \oplus (1)^{6}
\oplus (0)^{3}.
\label{decompose}
\end{eqnarray}
Here, the left hand side denote the direct
product of the four $S=1$ spins while the right hand side
denotes the decomposition in the total spin basis. 

The sum of any two
nearest neighbor spins on the lattice ($S_{pair}(S_{pair}+1) =
(\vec{S}_{i} + \vec{S}_{j})^{2}$) can only attain the three values
$S_{pair} = 0,1,2$. If any two spins forming a given plaquette are in
a singlet state ($S_{pair} =0$), then the total spin of the plaquette cannot
exceed two, $S_{\Box} \le 2$. It follows that if the polynomial
$\mathcal{P}$ of Eq.(\ref{hp.}) is non-negative definite and equal to
zero for $S_{\Box} = 0,1,2$, then any state $| \psi \rangle$ which is a
superposition of dimer coverings of the lattice each of which 
has (at least) one dimer per each plaquette then
\begin{eqnarray}
| \psi \rangle = \sum_{P} \alpha_{P} \prod_{ij \in P} |S_{ij} \rangle,
\label{dimer} 
\end{eqnarray}
(with such dimer coverings labeled by $P$) 
is a ground state of $H$. The above singlet state is given by
\begin{eqnarray}
| S_{ij} \rangle = \frac{1}{\sqrt{3}}[|m_{i} = 1, m_{j} = - 1 
\rangle \nonumber
\\ + |m_{i} = - 1, m_{j} = 1 \rangle - |m_{i} = 0, m_{j} = 0 \rangle].
\label{singlet.}
\end{eqnarray}
Putting all of the pieces together, any state of the form
Eqs.(\ref{dimer},\ref{singlet.}) is a ground state of the Hamiltonian
of Eqs. (\ref{h.}, \ref{hp.}) for any non-negative definite polynomial
$\mathcal{P}$ which vanishes for $S_{\Box} = 0,1,2$. For concreteness,
we now explicitly give an example of such a polynomial,
\begin{eqnarray}
h_{\Box} = \vec{S}_{\Box}^{2}(\vec{S}_{\Box}^{2} - 2)(\vec{S}_{\Box}^{2}- 6).
\label{explicit.}
\end{eqnarray}
Many other non-negative definite polynomials over all possible
plaquette spin values can be written down such that they vanish only
for $S_{\Box} =0,1,2$.  Any of the dimer coverings of the form given
by Eqs.(\ref{dimer},\ref{singlet.}) is a ground state of the
Hamiltonian given by Eqs.(\ref{h.}, \ref{explicit.}).

\subsection{Geometry of the ground states of the $S=1$ system} 

Enumerating all possible ground states of the form of Eq.(\ref{dimer})
amounts to the task of finding all possible dimer coverings of the
on a square lattice with periodic boundary conditions such that 
every plaquette hosts, at least, one dimer. 
The problem of hard core dimer
coverings was investigated by 
\cite{Fisher} wherein the number of
fully packed coverings on an infinite lattice was 
found to scale as $N \simeq \exp[G A/\pi]$ with $A$ the number of sites
in the lattice and $G$ Catalan's constant
($G = \sum_{n=0}^{\infty} \frac{(-1)^{n}}{(2n+1)^{2}} \approx 0.915965$). 
When the condition of (at least) a single dimer per plaquette
is imposed on a square lattice with periodic boundary 
conditions, we find that there are $2^{L}$ (with $L$
the lattice length) collinear 
diagonal coverings\cite{BT}. Along each diagonal, 
we may decide at our whim, to
orient the dimers either vertically or horizontally. As the number of
diagonals is $L$ and as there are two such choices
for dimer orientations (horizontal/vertical)
for each diagonal,
the number of individual dimer ground states scales as $2^{L}$. 
Apart from these $2^{L}$ ``stripe'' states, there are 
no other dimer coverings such that one dimer
appears in each plaquette.

Implementing these geometric constraints on the general 
ground states of the $S=1$ system of Eqs.(\ref{dimer},\ref{singlet.}),
the allowed ground states read
\begin{eqnarray}
| \psi \rangle = \sum_{P \pm } \alpha_{P \pm} \prod_{i \in A} 
|S_{i, 
i + \hat{\phi}(i_{x} \pm i_{y})} \rangle.
\label{dimer-star} 
\end{eqnarray}
The difference between the general dimer states of Eq.(\ref{dimer})
and Eq.(\ref{dimer-star}) lies in the explicit form of the allowed 
partitions $P$ which is discussed below.  
In Eq.(\ref{dimer-star}), 
the sites $i \in A$ belong to one sub-lattice (say the one given
by $(i_{x} \pm i_{y})$ being even with $i_{x,y}$ the x and y
components of $i$) and the unit vectors $\hat{\phi}(i_{x} \pm i_{y}) =
\hat{e}_{x}~\mbox{or} -\hat{e}_{y}$ are random Ising variables. The
configurations $P_{+}$ are those in which the orientation of the
dimers (horizontal or vertical) is fixed along right-tilting diagonals
($i_{x} - i_{y} = const$) whereas in $P_{-}$ are those in which the
orientation of the dimers is fixed along left-tilting diagonals (fixed
$i_{x} + i_{y}$).  As there are $2^{L}$ possible sequences in $P_{+}$
for the field $\hat{\phi}(i_{x} \pm i_{y})$ along all of the
right-tilting diagonals (whose number is $L$) with a similar number of
possible coverings $P_{-}$, the set of coverings $\{P_{+}, P_{-}\}$
spans ${\cal{O}}(2^{L+1})$ configurations. Similarly, additional dimer
coverings where the left/bottom most position of the dimer may span
both sub-lattices appear.
Various orders can be stabilized by augmenting the Hamiltonian
by an additional field favoring this or the other ground states
\cite{BT}.  A cartoon of such an external field of relative strength
$g$ stabilizing the parallel (all $\hat{\phi} = \hat{e}_{x}$ in
Eq.(\ref{dimer-star})) and zig-zag phase ($\hat{\phi} = \hat{e}_{x},
\hat{e}_{y}, \hat{e}_{x}, \hat{e}_{y}, ...$ uniformly on consecutive
parallel diagonals) at different values of the external field $g$ is
shown in Figure 1.  Such a term may be, e.g.,
\begin{eqnarray}
H_{g} =   [\sum_{i_{x}+ i_{y} \equiv 0 (~\mbox{mod}~ 4)} (\vec{S}_{i} 
+ \vec{S}_{i+ \hat{e}_{x}})^{2} \nonumber
\\ -  g  \sum_{i_{x}+ i_{y} \equiv 2 (~\mbox{mod}~ 4)} \{ (\vec{S}_{i} 
+ \vec{S}_{i+ \hat{e}_{y}})^{2} \nonumber
\\ - (\vec{S}_{i} 
+ \vec{S}_{i+ \hat{e}_{x}})^{2}\}].
\label{g.}
\end{eqnarray} 
Here, the first term of Eq.(\ref{g.}), 
spans sites $i$ belonging to half
of the sites of sub-lattice $A$ (the sites 
in which the sum of the
x and y components of the 
sites $\vec{i}$ is an integer multiple
of four, $i_{x} + i_{y} = 4m$ with $m$
an integer). The second term
in Eq.(\ref{g.}), 
spans sites which belong to 
the other half of
the sub-lattice $A$.

\begin{figure}[!htb]
  \includegraphics[angle=0,width=9cm]{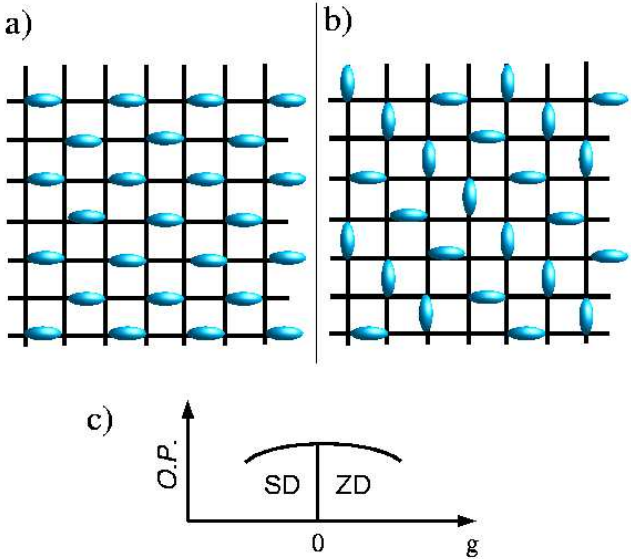}
  \vspace{-3.5cm}
  \caption{Reproduced from \cite{BT}- two possible dimer coverings of
    the square lattice (a, b) such that each plaquette
hosts one dimer.  In the context of the $S=1$
    Hamiltonian discussed here, the ovals denote a singlet state
    $|S_{ij} \rangle$ of Eq.(\ref{singlet.}) formed between two
    nearest neighbor sites $i$ and $j$.  Such coverings can be
    stabilized by augmenting the Hamiltonian of
    Eq.(\ref{h.},\ref{hp.}) with additional terms similar to those of
    Eq.(\ref{g.}).  As a function of $g$ a transition occurs between
    the single diagonal (SD) configuration of (a) and the zigzag (ZD)
    phase of (b). The shorthand O.P. denotes
an order parameter in each of the phases.}\label{fig}
\end{figure} 

\vspace{-0.5cm}

This model hosts a form of topological
order in some of its ground states 
which is seen by string correlators.
Any ground
state 
\begin{eqnarray}
| \psi_{P} \rangle = \prod_{ij \in P} |S_{ij} \rangle,
\label{dimer+} 
\end{eqnarray}
is a direct product of two leg ladders (which are oriented along 
the diagonals) which are
fully occupied by singlet
states along each rung. Correlators involving spins along
a line of consecutive singlets in $P$ alone 
display the same behavior
as that of a two leg ladder.
As the wave-function decouples into a product 
of two spin singlet wave-functions along each rung 
of the ladder, it is readily verified that in 
the states of Eq.(\ref{dimer+}), the string correlator
\begin{eqnarray}
\langle S_{1,i}^{z} S_{2,i}^{z} \exp[i \pi \sum_{k=i+1}^{j-1}
(S_{1k}^{z} + S_{2k}^{z})] S_{1,j}^{z} S_{2,j}^{z} \rangle = \frac{4}{9},
\label{topo1}
\end{eqnarray}
irrespective of the separation between rungs $i$ and $j$.
Here, the indices 1 and 2 denote the location of the spin
on one of the endpoints of the rung. 
A correlator of the form of Eq.(\ref{topo1}) was first
introduced by Todo et al. \cite{todo}.

\subsection{An extended $t-J_{z}$ $S=1/2$ fermionic model}

We now map our solvable $S=1$ Hamiltonians and their
ground states onto an electronic ($S=1/2$ fermion) problem. To this
end, we employ a Jordan-Wigner transformation mapping the three states
of a $S=1$ spin onto the three states of a constrained $S=1/2$ fermion
(one in which a forbidden double occupancy at
any given site is strictly enforced).

Rather explicitly, following \cite{BO}, the mapping between these
constrained $S=1/2$ fermions and the $S=1$ spin operators is
\begin{eqnarray}
S_{j}^{+} = \sqrt{2}(\overline{c}_{j \uparrow}^{\dagger} K_{j} +
K^{\dagger}_{j} \overline{c}_{j \downarrow} ), \nonumber\\
S_{j}^{-} = \sqrt{2}(K^{\dagger}_{j} \overline{c}_{j
\uparrow} + \overline{c}_{j \downarrow}^{\dagger} K_{j}),
\nonumber\\
S_{j}^{z} =  \overline{c}_{j \uparrow}^{\dagger}  
\overline{c}_{j \uparrow} -
\overline{c}_{j \downarrow}^{\dagger} \overline{c}_{j \downarrow}
\end{eqnarray}
Here the two dimensional kink operators leading to the correct
(fermionic) statistics upon transmutation via Aharonov-Bohm phases 
by flux attachment \cite{flux} are 
\begin{eqnarray}
K_{j} \equiv \exp[i \sum_{k} \theta_{k,j} \overline{n}_{k}]
\end{eqnarray}
with $ \overline{n}_{k} = \overline{c}_{k \uparrow}^{\dagger}  
\overline{c}_{ k \uparrow} +
\overline{c}_{k \downarrow}^{\dagger} \overline{c}_{k \downarrow}$
the total occupancy of site $k$ and $ \theta_{k,j}$ the 
angle between the line connecting sites $k$ and $j$
and, say, the horizontal x-axis. Here the sum
spans all lattice sites $k$. 

Explicitly inserting this into the scalar products forming $h_{\Box}$
in Eqs.(\ref{h.}, \ref{explicit.}), we find that in $h_{\Box}$ any
former appearance of $(\vec{S}_{a} \cdot \vec{S}_{b})$ in the $S=1$
problem is now replaced by a $t-J_{z}$ like pair term augmented by 
superconducting pieces with phases in tow,
\begin{eqnarray}
T_{ab} \equiv 4 s_{a}^{z} s_{b}^{z} + (\overline{c}_{a \uparrow}^{\dagger}
\overline{c}_{b \uparrow} e^{i \phi_{ab}}
+ \overline{c}_{a \uparrow}^{\dagger} 
\overline{c}_{b \downarrow}^{\dagger} e^{i \overline{\phi}_{ab}} \nonumber
\\ + \overline{c}_{a \downarrow} \overline{c}_{b \uparrow} 
e^{-i \overline{\phi}_{ab}} + \overline{c}_{a \downarrow} 
\overline{c}_{b \downarrow}^{\dagger} e^{- i \phi_{ab}} + h.c.).
\label{T.}
\end{eqnarray}
Here, $\vec{s}_{a,b}$ denote the $S=1/2$ spin operators 
of the constrained fermions on sites $a$ and $b$ which 
belong to the same plaquette, and the phases 
\begin{eqnarray}
\phi_{ab} \equiv \sum_{c} [(\theta_{ac} 
- \theta_{bc}) \overline{n}_{c}], \nonumber
\\ \overline{\phi}_{ab} \equiv \sum_{c} [(\theta_{ac} 
+ \theta_{bc}) \overline{n}_{c}]
\label{phi.}
\end{eqnarray}
with the sum performed over all lattice sites $c$.
The operator $T_{ab}$ is an extended variant of the
nearest neighbor interaction appearing in Eq.(\ref{tjzHam})
with $J_{z}=4t$ having additional terms (both additive
and multiplicative). 

With this substitution any appearance of $\vec{S}_{\Box}^{2}$ in the
$S=1$ problem is simply replaced by
\begin{eqnarray}
\vec{S}_{\Box}^{2} \to 2 (\sum_{(a b)} T_{ab} +  4)
\end{eqnarray}
where $(ab)$ span all six pairs of sites which may be chosen from the
four sites comprising the plaquette. The Hamiltonian governing the
$S=1/2$ fermionic system is
\begin{eqnarray}
H =  \sum_{\Box}[(\sum_{(a b) \in \Box} T_{ab} +1) \nonumber
\\ \times
(\sum_{(cd) \in \Box } T_{cd} + 3)  
(\sum_{(fg) \in \Box} T_{fg} + 4)].
\label{Hfinal}
\end{eqnarray} 
This is a particular variant of an extended $t-J_{z}$ model
with fixed couplings. 

Similarly, the electronic state ground
state of Eq.(\ref{Hfinal}) (derived from
$|S_{ij} \rangle$) is
\begin{eqnarray}
|\chi_{ij} \rangle = \frac{1}{\sqrt{3}}(|\uparrow_{i} \downarrow_{j} \rangle
+ | \downarrow_{i} \uparrow_{j} \rangle - | 0_{i} 0_{j} \rangle).
\end{eqnarray}
Here, $|\uparrow_{i} \rangle (|\downarrow_{i} \rangle)$
denotes a single electron at site $i$ with spin up (down)
while $|0_{i} \rangle$ corresponds to an empty site
at site $i$. Thus a ground state of a Hamiltonian given
by Eq.(\ref{Hfinal}) with the definition of Eq.(\ref{T.}) is
\begin{eqnarray}
|\psi \rangle = \sum_{P} \alpha_{P} \prod_{ij \in P}|\chi_{ij} \rangle.
\label{dimer-star-star}
\end{eqnarray}
In $|\psi \rangle$, at all lattice sites $k$, the occupancy $\langle
n_{k} \rangle = 2/3$ and the phases $\phi_{ab}, \overline{\phi}_{ab}$
vanish if we replace, in Eq.(\ref{phi.}), $\overline{n}_{k}$ by
$\langle \overline{n}_{k} \rangle$.  Eq.(\ref{dimer-star-star}) is
superposition of states with different particle numbers; this is a
consequence of the superconducting pair creation/ annihilation terms
in Eq.(\ref{T.}). The number of dimer coverings (the dimer 
partitioned denoted by $P$ an example of which 
is shown in Fig.(\ref{fig})) is exponential in the perimeter
of the system

All stated above is exact. In the following paragraph (and here alone)
we depart from the rigorous results derived above 
and briefly speculate about possible
extensions. As within the
ground state of Eq.(\ref{dimer-star-star}) there is no kinetic motion
of the electrons- we might speculate that the transmutation statistics
might well become irrelevant (as no transmutations occur). If this is
the case, $| \psi \rangle $ of Eq.(\ref{dimer-star-star}), will be an
eigen-state of Eq.(\ref{Hfinal}) with all phases $\phi_{ab},
\overline{\phi}_{ab}$ set to zero.

\subsection{Hard Core Bose $S=1/2$ extended $t-J_{z}$ models}

If we replace the hard core $S=1/2$ fermions by
hard core particles obeying other statistics, we find a
generalization of the above. Rather explicitly, if hard core particles
given by $\overline{b}_{\alpha, \sigma}, 
\overline{b}_{\alpha, \sigma}^{\dagger}$, with $\sigma = \uparrow, 
\downarrow$, obey
the fractional statistics of a phase change of $e^{i \pi \alpha}$
under the interchange of two particles (e.g. $\alpha = 0$ is the hard
core bosonic case) then Eq.(\ref{dimer-star-star}) is a ground state
of Eq.(\ref{Hfinal}) subject to the substitution $T_{ab} \to
T_{ab}^{\alpha}$ with
\begin{eqnarray}
T_{ab}^{\alpha} \equiv 4 s_{a}^{z} s_{b}^{z} + 
(\overline{c}_{a \downarrow}^{\dagger}
\overline{c}_{b \uparrow} e^{i \alpha \phi_{ab}}
+ \overline{c}_{a \uparrow}^{\dagger} 
\overline{c}_{b \downarrow}^{\dagger} 
e^{i \alpha \overline{\phi}_{ab}} \nonumber
\\ + \overline{c}_{a \downarrow} \overline{c}_{b \uparrow} 
e^{-i \alpha \overline{\phi}_{ab}} + \overline{c}_{a \downarrow} 
\overline{c}_{b \downarrow}^{\dagger} e^{- i \alpha \phi_{ab}} + h.c.).
\label{T..}
\end{eqnarray}
In the case of spin $S=1/2$ hard core bosons
[by which we refer to two flavor ($\sigma = \uparrow, \downarrow$)
hard core bosons],   
the resulting Hamiltonian explicitly 
reads
\begin{eqnarray}
H =  \sum_{\Box}[(\sum_{(a b) \in \Box} 4 s_{a}^{z} s_{b}^{z} + 
(\overline{c}_{a \downarrow}^{\dagger}
\overline{c}_{b \uparrow}
+ \overline{c}_{a \uparrow}^{\dagger} 
\overline{c}_{b \downarrow}^{\dagger}  \nonumber
\\ + \overline{c}_{a \downarrow}   \overline{c}_{b \uparrow} 
 + \overline{c}_{a \downarrow} 
\overline{c}_{b \downarrow}^{\dagger} + h.c.) +1) \nonumber
\\ \times
(\sum_{(cd) \in \Box }4 s_{c}^{z} s_{d}^{z} + 
(\overline{c}_{c \downarrow}^{\dagger}
\overline{c}_{d \uparrow}
+ \overline{c}_{c \uparrow}^{\dagger} 
\overline{c}_{d \downarrow}^{\dagger} \nonumber
\\ + \overline{c}_{c \downarrow} \overline{c}_{d \uparrow} 
 + \overline{c}_{c \downarrow} 
\overline{c}_{d \downarrow}^{\dagger}  + h.c.)  + 3)  \nonumber
\\ \times (\sum_{(fg) \in \Box}4  s_{f}^{z} s_{g}^{z} + 
(\overline{c}_{f \downarrow}^{\dagger}
\overline{c}_{g \uparrow} 
+ \overline{c}_{f \uparrow}^{\dagger} 
\overline{c}_{g \downarrow}^{\dagger}  \nonumber
\\ + \overline{c}_{f \downarrow} \overline{c}_{g \uparrow} 
+ \overline{c}_{f \downarrow} 
\overline{c}_{g \downarrow}^{\dagger}  + h.c.) + 4)].
\label{Hfinal*}
\end{eqnarray} 
The states given by Eq.(\ref{dimer-star-star}) are 
exact ground states of this short-range Hamiltonian. 
The frustrated kinetic hopping
parameters and competing nearest neighbor and next nearest
neighbor antiferromagnetic interactions physically adhere to the
resulting ground state. The precise form of the diagonal formation of
singlet dimers and high degeneracy of these states are non-trivial.
Similar degeneracies appear in many other systems \cite{BN}
where effective dimensional reduction occurs. 
By extending the correlator of Eq.(\ref{topo1}) 
to the $S=1/2$ arena
by means of the 
transformations of \cite{BO} and noting that
$\exp[2 \pi i s^{z}_{i} ] = \exp[i \pi \overline{n}_{i}]$
at all lattice sites $i$, we
find a counterpart
to the Ogata-Shiba correlators \cite{og}
for two leg ladders
\begin{eqnarray}
\langle \tilde{\chi}_{P} | s^{z}_{1,i} s^{z}_{2,i} \nonumber
\\  
\exp[i \pi \sum_{k=i+1}^{j-1}( \overline{n}_{1,k} + \overline{n}_{2,k})]
s^{z}_{1,j} s^{z}_{2,j} | \tilde{\chi}_{P} \rangle  = \frac{1}{36}.
\end{eqnarray}
Here, $| \tilde{\chi}_{P} \rangle = \prod_{ij \in P} | \chi_{ij} \rangle$.

Other ground states
can easily be proven 
along the same lines. As the fully ferromagnetic 
states $S_{i}^{z} = +1$ for all $i$
or $S_{i}^{z} = -1$ are ground
states of the spin $S=1$ Hamiltonian 
\begin{eqnarray}
H = - \sum_{\Box} \vec{S}_{\Box}^{2},
\end{eqnarray}
the maximally occupied 
(by hard core bosons) fully polarized ferromagnetic
states $s_{i}^{z} = 1/2$ ($n_{i} =1$) for all $i$
or $s_{i}^{z} = -1/2$ ($n_{i} =1$) at all sites $i$
are ground states of the $s=1/2$ Bose extended $t-J_{z}$ model
\begin{eqnarray}
H = - \sum_{\Box} 
[\sum_{(a b) \in \Box} 4 s_{a}^{z} s_{b}^{z} + 
(\overline{c}_{a \downarrow}^{\dagger}
\overline{c}_{b \uparrow}
+ \overline{c}_{a \uparrow}^{\dagger} 
\overline{c}_{b \downarrow}^{\dagger} 
 \nonumber
\\ + \overline{c}_{a \downarrow} \overline{c}_{b \uparrow} 
 + \overline{c}_{a \downarrow} 
\overline{c}_{b \downarrow}^{\dagger} + h.c.)].
\end{eqnarray}

\section{Other Lattices}

\begin{figure}[t!]
\vspace*{-0.5cm}
\includegraphics[angle=0,width=9cm]{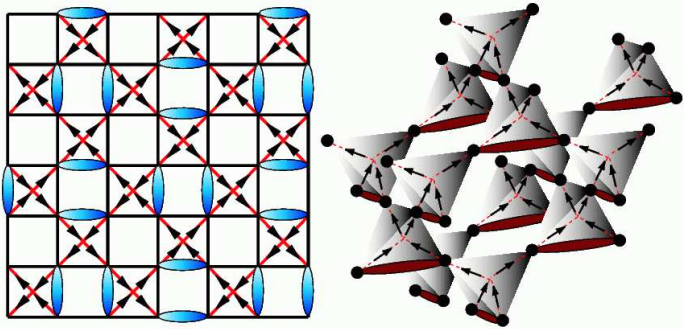}
\vspace{-1.5cm}
\caption{ (Color online.) 
From \cite{NBNT}. 
Highly regular ground states on the checkerboard and pyrochlore 
lattices. The ovals denote singlet dimer states. The arrows denote the 
representations of these dimer states within the six--vertex model (see 
text). On each plaquette (tetrahedron) the dimer connects the bases of 
the two incoming arrows.}
\label{PR}
\bigskip 
\end{figure}

We may repeat our derivations above to derive the ground states
on any other lattice in which four fermions
in any given basic unit interact with each other according to
Eq.(\ref{Hfinal}). One such example is afforded by 
the three dimensional pyrochlore lattice and its 
dimensional checkerboard rendition. The pyrochlore lattice is 
composed of corner sharing tetrahedra such that each vertex is common
to two 4-site units (see the right hand side of Fig.(\ref{PR}). The geometry 
is rather common in transition metal and rare earth oxide
systems, specifically in the pyrochlore (the origin of
the name) and spinel structures. The checkerboard lattice (see
the left hand side of Fig.(\ref{PR})) is a 2d projection of the 
pyrochlore structure: the crossed plaquettes in the left
hand side of Fig.(\ref{PR}) represent 
the basic tetrahedral units.

On the pyrochlore/checkerboard 
lattices the open structure of tetrahedra/crossed plaquettes 
creates, by comparison to the square lattice, a less constrained 
system, and the number of dimer configurations in the ground--state manifold 
is strongly enhanced, the still more massive degeneracy implying similarly 
exotic physics in this case. 

In addition to the simple coverings shown in Fig.~{\ref{PR}}, a far richer 
variety of states exists. A detailed study of 
dimer states on the pyrochlore and checkerboard lattices
is provided in \cite{NBNT}. 
A mapping onto the spin--ice problem 
provides a useful 
classification of these ground states. Here, 
the states with one dimer per tetrahedral unit defined by 
Eq.~(\ref{dimer}) map exactly to the spin--ice problem \cite{gingras, 
anderson} where, motivated by the structure of H atoms in solid water 
\cite{fowler,Pauling}, two of 
the sites of any elementary unit (a tetrahedron) are associated with an 
ingoing arrow pointing towards the center of the tetrahedron and two sites 
lie on arrows pointing outwards. We label the two sites belonging to a 
singlet dimer $|S_{ij} \rangle$ in a given tetrahedron $\Box$ by two 
incoming arrows from sites $i$ and $j$ to the center of the unit $\Box 
\equiv ijkl$. In this fashion it is clear that the system is mapped to 
a set of continuous directed lines such that each tetrahedral unit has 
exactly two incoming and two outgoing arrows relative to its center. 
An example of this mapping is illustrated in Fig.~\ref{PR}. The mapping 
is one--to--one: any spin--ice configuration determines a unique 
singlet--covering state with one dimer per tetrahedral unit and vice 
versa. The algebraic correlations in the spin-ice model
dictate an algebraic decay of correlations in
the $d=2,3$ dimensional $S=1/2$ Fermi problems 
of Eq.(\ref{Hfinal}) on the checkerboard and pyrochlore
lattices : $\langle T_{ab} T_{mn} \rangle \sim |r|^{-d}$
for large separations $|r|$. Here, $|r|$ denotes
the distance between the two 
site pairs ($(ab)$ and $(mn$)) \cite{NBNT}, \cite{Baxter}, \cite{Lieb}.

\section{Conclusion} 

In conclusion, we constructed spin $S=1/2$ extended
$t-J_{z}$ models (both fermionic and hard core bosonic) which harbor
easily provable exact ground states.  We illustrated how this method
can be used to engineer non-uniform diagonal dimer state and other
configurations. Applying similar ideas, many other
solvable systems can be constructed on other lattices.  

This work was partially sponsored by the
Swedish Research Council.


\begin{thebibliography}{99}
\bibitem{explain_qe} "Quasi-exactly solvable" here means that
we can exactly determine all ground states 
yet do not solve the finite temperature problem.
\end{thebibliography}
\end{document}